# Image Steganography Method Based on Brightness Adjustment

Youssef Bassil

LACSC – Lebanese Association for Computational Sciences
Registered under No. 957, 2011, Beirut, Lebanon

Email: youssef.bassil@lacsc.org

**Abstract** – Steganography is an information hiding technique in which secret data are secured by covering them into a computer carrier file without damaging the file or changing its size. The difference between steganography and cryptography is that steganography is a stealthy method of communication that only the communicating parties are aware of; while, cryptography is an overt method of communication that anyone is aware of, despite its payload is scribbled. Typically, an irrecoverable steganography algorithm is the algorithm that makes it hard for malicious third parties to discover how it works and how to recover the secret data out of the carrier file. One popular way to achieve irrecoverability is to digitally process the carrier file after hiding the secret data into it. However, such process is irreversible as it would destroy the concealed data. This paper proposes a new image steganography method for textual data, as well as for any form of digital data, based on adjusting the brightness of the carrier image after covering the secret data into it. The algorithm used is parameterized as it can be configured using three different parameters defined by the communicating parties. They include the amount of brightness to apply on the carrier image after the completion of the covering process, the color channels whose brightness should be adjusted, and the bytes that should carry in the secret data. The novelty of the proposed method is that it embeds bits of the secret data into the three LSBs of the bytes that compose the carrier image in such a way that does not destroy the secret data when restoring back the original brightness of the carrier image. The simulation conducted proved that the proposed algorithm is valid and correct. As future work, other image processing techniques are to be examined such as adjusting the contrast or the gamma level of the carrier image, enabling the communicating parties to more flexibly configure their secret communication.

## 1. Introduction

The past years have seen a growing interest in data confidentiality to defend against eavesdropping and unauthorized access to digital properties. This has led to the development of endless methods and techniques in the field of information security, two of which are mainly Cryptography and Steganography. Basically, the former provides data security by distorting the secret message in a way that no one, except the communicating parties, can understand it; whereas, the latter provides data security by hiding the very existence of the secret message in a way that no one, except the communicating parties, can know about its presence [1]. Technically speaking, cryptography converts the secret data into some other garbage form of data called encrypted data, which are then communicated overtly between the sender and the receiver. On the other hand, steganography embeds the secret data into what is known as a carrier file, making the data totally imperceptible to any party including the sender and receiver themselves, though they know the exact location of the secret data in the carrier file [2]. In effect, the advantage of steganography over cryptography is that although in cryptography the transmitted secret data cannot be read by unauthorized third parties, they can still draw attention and reveal the fact that a secret communication is taking place. In contrast, steganography is stealthy as it conveys the secret data through an innocent normal-looking carrier file, avoiding arousing an eavesdropper's suspicion. Steganography has two foremost requirements: The first one is imperceptibility of the carrier file which refers to totally obscuring the secret data in the carrier file without damaging it or changing its original size [3]. The second one is irrecoverability of the secret data which refers to preventing eavesdroppers from recovering the secret data by reverse-engineering the steganography algorithm. In image steganography, one way to promote irrecoverability is to digitally process the carrier image after concealing the secret data into it, for instance, flipping, rotating, mirroring, adjusting its contrast or even its brightness. Consequently, if by any means the carrier image is inspected by eavesdroppers, the contained secret data would be already transformed and no one can tell that they even existed. However, this comes with a price as digitally processing the carrier image is often an irreversible process [4] which would destroy the secret data hidden inside the carrier image and prevent them from being recovered by the concerned parties.

This paper proposes a new image steganography method based on adjusting the brightness of the carrier image after covering the secret data into it. The method is geared by a parameterized algorithm which allows the communicating parties to specify 1) the amount of brightness to apply on the carrier image after the completion of the covering process; 2) the color channels whose brightness should be adjusted; and 3) the bytes that should house the secret data. The proposed method is designed for textual data as well as for any form of data as long as they can be converted into a binary format. The secret data are covered in the three least significant bits (LSB) of the bytes composing the carrier image. However, these bytes are not selected sequentially as in traditional LSB steganography techniques; rather, they are selected based on their intensities in such a way that does not destroy the secret data when restoring back the original brightness of the carrier image. Fundamentally, the fact of transforming the carrier image into a brighter version shadows the hidden secret data, preventing their recovery by unauthorized parties.

## 2. Image Steganography and Its Applications

In essence, image steganography is drawn upon the visual limited capabilities of the human visual system (HVS) which cannot detect the slight intensity variation of the pixels of an image [5]. As a result, hiding the secret data into the LSBs of the pixels composing the carrier image would change marginally the color intensities of the image so much so that it would be unnoticeable by a human naked eye. Thus, an unauthorized observer cannot distinguish between the original carrier image and the tampered one, i.e., the one that carries the secret data into the LSBs of its pixels.

The applications of image steganography are diverse; they include secret communication, digital watermarking, and data integrity [6, 7].

**Secret Communication:** Characteristically, transmitting a cryptographic message may raise unwanted suspicions as it travels overtly. Besides, cryptography is restricted by law in many situations. Contrariwise, steganography does not publicize secret communication; and hence, it evades the message from being detected and recovered by malicious parties.

**Digital Watermarking:** A secret copyright mark also known as digital watermark can be embedded inside an image to verify its authenticity. A digital watermark can be used to identify the owner of an intellectual property such as an image, audio, or video file. It also helps fight against copyright infringements as it detects the source of illegally copied materials.

**Data Integrity:** It refers to the assertion of data that they are accurate after they have been transmitted over the Internet and received by the recipient. Using steganography, a special mark can be embedded within an image file to determine that no variations, compromises, or damages have occurred to the image after it is received by the recipient. This form of steganography is referred to as Digitally Signing images so as to be able to confirm their reliability at any time.

## 3. State-of-the-Art in Image Steganography

So far, massive research work has been conducted in the development of steganography for digital images. One of the earliest techniques is the LSB technique which obscures data communication by inserting the secret data into the insignificant parts of the pixels of an image file, more particularly, into the least significant bits (LSB) [8]. The modified version of the image, which is called carrier file or stego file, is then sent to the receiver through a public channel. The foremost requirement of the LSB technique is that it should not exhibit any visual signs in the carrier image so as to not give any indications that secret data are being communicated covertly.

Basically, the LSB technique is an insertion-based image steganography method that embeds secret data into uncompressed computer image files such as BMP and TIFF. In this technique, the data to hide are first converted into a series of bytes, then into a series of smaller chunks each of which is of size *n* bits. Then, *n* LSBs of the pixels of the carrier image are replaced by each of the chunks of the original data to hide. The ultimate result of this operation is a carrier image carrying the secret data into the LSBs of its pixels. As the color values that are determined by LSBs are insignificant to the naked eye, it is hard to tell the difference between the original image and the tampered one, taking into consideration that no more than a certain number of LSBs were used to conceal the secret data; Otherwise, visual artifacts and damages would be produced in the carrier image which would in turn draw suspicions and raise attention about something unusual in the carrier image. For instance, in 24-bit True Color BMP images, using more than three LSBs per color component to hide data may result in perceptible artifacts in the carrier image [9]. As an illustration for the LSB technique, let's say that the letter H needs to be hidden into an 8-bit grayscale bitmap image. The ASCII representation for letter H is 72 in decimal or 01001000 in binary. Assuming that the letter H is divided into four chunks each of 2 bits, then four pixels are needed to totally hide the letter H. Moreover, assuming that four consecutive pixels are selected from the original image whose grayscale values are denoted by $P_1$=11011000, $P_2$=00110110, $P_3$=11001111, and $P_4$=10100011, then substituting every two LSBs in every of these four pixels by a 2-bit chunk of the letter H, would result in a new set of pixels denoted by $P_1$= 110110**01**, $P_2$=001101**00**, $P_3$=110011**10**, and $P_4$=101000**00**. Despite changing the actual grayscale values of the pixels, this has little impact on the visual appearance of the carrier image because characteristically, the Human Visual System (HVS) cannot differentiate between two images whose color values in the high frequency spectrum are marginally unalike [10].

On the other hand, other steganography techniques and algorithms for digital images have been proposed and researched both in spatial and frequency domains. They include masking and filtering [11], encrypt and scatter [12], transformation [13], and BPCS [14] techniques.

**Masking and Filtering Technique:** This technique is based on digital watermarking but instead of increasing too much the luminance of the masked area to create the digital watermark, a small increase of luminance is applied to the masked area making it unnoticeable and undetected by the naked eye. As a result, the lesser the luminance alteration, little the chance the secret message can be detected. Masking and filtering technique embeds data in significant areas of the image so that the concealed message is more integral to the carrier file.

**Encrypt and Scatter Technique:** This technique attempts to emulate what is known by White Noise Storm which is a combination of spread spectrum and frequency hopping practices. Its principle is so simple; it scatters the message to hide over an image within a random number defined by a window size and several data channels. It uses eight channels each of which represents 1 bit; and consequently, each image window can hold 1 byte of data and a set of other useless bits. These channels can perform bit permutation using rotation and swapping operations such as rotating 1 bit to the left or swapping the bit in position 3 with the bit in position 6. The niche of this approach is that even if the bits are extracted, they will look garbage unless the permutation algorithm is first discovered. Additionally, the encrypt and scatter technique employs DES encryption to cipher the message before being scattered and hidden in the carrier file.

**Transformation Technique**: This technique is often used in the lossy compression domain, for instance, with JPG digital images. In fact, JPG images use the discrete cosine transform (DCT) to perform compression. As the cosine values cannot be calculated accurately, the DCT yields to a lossy compression. The transformation-based steganography algorithms first compress the secret message to hide using DCT and then integrate it within the JPG image. That way, the secret message would be integral to the image and would be hard to be decoded unless the image is first decompressed and the location of the hidden message is recovered.

**BPCS Technique**: This technique which stands for Bit-Plane Complexity Segmentation Steganography, is based on a special characteristic of the Human Visual System (HVS). Basically, the HVS cannot perceive a too complicated visual pattern as a coherent shape. For example, on a flat homogenous wooden pavement, all floor tiles look the same. They visually just appear as a paved wooden surface, without any indication of shape. However, if someone looks closely, every collection of tiles exhibits different shapes due to the particles that make up the wooden tile. Such types of images are called vessel images. BPCS Steganography makes use of this characteristic by substituting complex regions on the bit-planes of a particular vessel image with data patterns from the secret data.

## 4. Proposed Solution

This paper proposes a new image steganography method for hiding digital data, regardless of their types - whether text, files, or documents, into uncompressed image files. The method is based on adjusting the brightness of the carrier image after covering the secret data into it. The algorithm used is parameterized, in that, it enables the communicating parties to decide on certain parameters prior to starting their secret communication. The first parameter specifies the amount (level) of brightness to apply on the carrier image after hiding the secret data into it. The second parameter specifies the color channels whose brightness is to be adjusted. Possible options are the Red channel alone, the Green channel alone, the Blue channel alone, the three channels altogether (overall image), or even a combination of them. The third parameter specifies, based on their intensities, the bytes into which the secret data are to be covered. These parameters are stored in predefined locations in the carrier image after adjusting the carrier's brightness. The secret data are covered in the three least significant bits (LSB) of the bytes that make up the carrier image. However, the selection of these bytes is not done sequentially as in traditional LSB steganography techniques; rather, it is done based on the third parameter which specifies the carrier bytes based on their intensities. The reason for this strategy is to avoid the destruction of the hidden secret data after restoring back the original brightness of the carrier image. Reverting back to the initial brightness level of the carrier image is necessary to recover the secret data.

### 4.1 Digital Image Processing – Adjusting Brightness

Digital image processing is the use of computer algorithms to process digital images in such a way that changes the image's properties including color, shape, orientation, sharpness, and brightness, among others [15]. In particular, brightness refers to the luminosity of an image. Adjusting the brightness of an image lightens or darkens all its colors equally by shifting its pixel intensity values up or down a tonal range. Mathematically, to increase the brightness of an image, a value must be added to the bytes that make up the image [16]. Unfortunately, increasing the brightness of an image can be sometimes irreversible as the resulting bytes may go above 255 which is the maximum value 8 bits can represent. As a result, bytes whose original values were 255 before increasing the image brightness would be equal to the bytes whose values became 255 after increasing the image brightness. Hence, any reversal process would make the original image loses some of its brightness information. Likewise, the opposite is true for negative values when byte intensities fall below 0 after decreasing the brightness of the image. Below is a sample code written using C#; its function is to adjust the brightness of an image by increasing it by an amount of 30.

```csharp
private void IncreaseBrightness (int amount )
{
    Bitmap bitmap = new Bitmap("c:\img.bmp");

    int red, green, blue;

    for (int y = 0; y < bitmap.Height; y++)
```

```
  {
    for (int x = 0; x < bitmap.Width; x++)
    {
      Color c = bitmap.GetPixel(x, y);

      red = (c.R + amount > 255) ? 255 : (c.R + amount);
      green = (c.G + amount > 255) ? 255 : (c.G + amount);
      blue = (c.B + amount > 255) ? 255 : (c.B + amount);

      bitmap.SetPixel(x, y, Color.FromArgb(red, green, blue));
    }
  }
}
```

## 4.2 Mathematical Conception of the Proposed Method

Essentially, adjusting the brightness of a digital image has many challenges, especially when employed in the context of steganography. This section tackles these challenges, paving the way for the mathematical conception of the proposed method.

**Irreversibility**: Basically, to increase the brightness of an image, the values of its composing bytes must be increased by some integer value. However, since 255 is the maximum value a byte can represent, increasing the image brightness may result in some byte values above 255. For instance, increasing the brightness of an image by an amount of 1 would make all bytes whose values are 255 to go above 255. Likewise, increasing the brightness by an amount of 40, would make all bytes whose values are 216 to go above 255. To remedy this problem, all pixels whose values went over 255 after adjusting the brightness of the image should be set to 255. As a result, all pixels whose original values before increasing the brightness of the image were 255 would be equal to all pixels whose values became 255 after increasing the brightness of the image. Therefore, one cannot determine the original value of a byte whose value became 255 after increasing the brightness of the image. For this reason, adjusting the brightness of an image is often an irreversible process. More importantly, the reversal operation can have very little impact on the human eye as the HVS (Human Visual System) cannot differentiate between slight variations in color intensities; however, it has a damaging impact on steganography as the hidden secret information would be partially destroyed.

A simple solution for this problem is not to hide the secret data in the bytes whose values would become greater than 255 after increasing the brightness of the image. For instance, if the amount of brightness to be applied is +10, the secret data should not be stored in the bytes whose values including their tampered LSBs are greater than 244 such as:

$$COVER(byte[i], secret\_data[j]) \text{ IF } COVER(byte[i], secret\_data[j]) < 255\text{-}brightness\_level. \quad (1)$$

The reason for taking into consideration the tampered LSBs as part of the byte values is because the secret data may, in some cases, increase these values to greater than *255-brightness_level*. Although, this method sounds plausible, it is in fact not, as digital images are sparse in high-intensity colors. This means that the secret data would be concealed more likely in a sequential manner in the carrier image, making them easier to be detected and recovered by eavesdroppers. A statistic conducted on 25 images, showed that on average only 3791 out 2,359,350 bytes have their values above 244.

**Sparsity of high-intensity colors**: Using the previously conceived solution (1), the secret data will end up being stored sequentially in the carrier image because digital images have very few bytes whose values are within the spectrum of high-intensity colors. A far more stealthy and effective solution would be to conceal the secret data into any byte, not necessarily high-intensity ones. An *upperbound_intensity* parameter is introduced to indicate the range of intensities from which the carrier bytes should be selected to store the secret data. This solution can be mathematically defined as follows:

For the covering process:

1. *byte[i]* = *upperbound_intensity* IF *byte[i]*>= *lowerbound_intensity* AND *byte[i]* <= *upperbound_intensity* , WHERE *upperbound_intensity+brightness_level<255* AND *lowerbound_intensity= SETnLSBs(upperbound_intensity , 0)*
2. **COVER**(*byte[i]*, *secret_data[j]*) IF *byte[i]* < *lowerbound_intensity*

For the uncovering process:

3. **UNCOVER**(*byte[i]- brightness_level*) IF *byte[i] - brightness_level < upperbound_intensity*

In the covering process, statement (1) ensures that the bytes bearing no secret data would have values greater or equal to *upperbound_intensity+brightness_level* after increasing the brightness of the image; while, statement (2) ensures that the secret data are stored in bytes whose values would be less than *upperbound_intensity+ brightness_level* after increasing the brightness of the image. Accordingly, secret data can be reliably uncovered using statement (3) by looking for all bytes whose values are less than *upperbound_intensity* after decreasing the brightness of the image.

As a result, the proposed method can conceal data in any byte of the image regardless of its intensity; thus, solving the problem of sparsity of high-intensities bytes by randomly scattering secret data over the carrier image and not storing them sequentially. This is not to mention that the proposed method is reversible despite adjusting the brightness of the carrier image. Using a pseudo-code, the proposed method can be expressed as follows:

```
// predefined or user selection
brightness_level = 20

// predefined → the secret data will be hidden in the 3 LSBs
// of every byte
n = 3

// predefined or user selection
upperbound_intensity = 100
```

```
// set the n LSBs of upperbound_intensity to 0 → 100 having
// its n=3 LSBs=0 is 96
lowerbound_intensity = SetnLSBs(upperbound_intensity, n, 0)

// 1. Preprocessing the carrier image
for_all_bytes_in_the_carrier_image DO:
if (byte[i] >= lowerbound_intensity && byte[i] <= upperbound_intensity)
{
    byte[i] = upperbound_intensity
}

// 2. The covering process
for_all_bytes_in_the_carrier_image DO:
if (byte[i] < lowerbound_intensity)
{
    Cover(byte[i] , secret_data[j])
}

// 3. Increasing brightness by brightness_level

// 4. The uncovering process
for_all_bytes_in_the_carrier_image DO:
if (byte[i] - brightness_level < upperbound_intensity)
{
    secret_data[j] = Uncover(byte[i] - brightness_level)
}
```

### 4.3 The Proposed Algorithm

The proposed algorithm comprises several steps to be executed in order to cover secret data into the bytes of a digital image. The steps are as follows:

1. The sender has to specify three parameters: The first parameter, denoted by *brightness_level*, is the brightness level to be applied on the carrier image after covering the secret data into it. The second parameter is the color channels whose brightness should be adjusted. It is denoted by *brightness_mode* and it has five different values: 1 indicating that only the Red channel should have its brightness adjusted, 2 indicating that only the Green channel should have its brightness adjusted, 3 indicating that only the Blue channel should have its brightness adjusted, 4 indicating that both the Red and Green channels should have its brightness adjusted, 5 indicating that both the Red and Blue channels should have their brightness adjusted, 6 indicating that both the Blue and Green channels should have their brightness adjusted, and 7 indicating that the Red, Green, and Blue channels should have their brightness adjusted. The third parameter, denoted by *upperbound_intensity*, is the range of intensities from which the carrier bytes are to be selected.
2. The secret data are converted into a binary form such as M={$m_0, m_1, m_2, ..., m_{n-1}$} where $m$ is a bit composing the secret data, and $n$ is the total number of bits.
3. The carrier image is preprocessed using the following equation: *byte[i]* = *upperbound_intensity* IF *byte[i]* >= *lowerbound_intensity* AND *byte[i]* <= *upperbound_intensity*, WHERE *upperbound_intensity+brightness_level<255* AND *lowerbound_intensity*= *SETnLSBs(upperbound_intensity* , 0)
4. A carrier byte is selected from the carrier image to store bits of the secret data. It is denoted by *byte[i]*; it is not selected sequentially but such as *byte[i] < lowerbound_intensity* where the step of *i* depends on the second parameter *brightness_mode*. It is worth noting that *lowerbound_intensity* is derived from the third parameter *upperbound_intensity* such as *lowerbound_intensity*= *SETnLSBs(upperbound_intensity* , 0)
5. The three LSBs of *byte[i]* are substituted by the three bits of the secret message M such as *byte[i]*={ $b_{i0}, b_{i1}, b_{i2}, b_{i3}, b_{i4}, m_j, m_{j+1}, m_{j+2}$ } where *byte[i]* is the i[th] carrier byte into which secret data M are to be hidden, $b$ is a bit from *byte[i]*, and $m_j$ is the j[th] bit from the secret data M. This step is repeated until all *m* are exhausted; thus completing the covering process.
6. The brightness of the carrier image is increased by adding to its bytes a value equal to the first parameter *brightness_level*. In fact, not all bytes are affected, only the ones that are part of the color channels indicated by the second parameter *brightness_mode*.
7. The three parameters *brightness_level* , *brightness_mode,* and *upperbound_intensity* that were already specified by the sender have to be communicated with the receiver prior to starting the secret communication. In effect, many solutions are possible, one of which is sending them via email, or handing them over the phone, or injecting them at the end of the carrier image, or embedding them into some predefined pixels locations in the carrier image using the traditional LSB technique.
8. Finally, the carrier image is sent to the receiver.

As for the uncovering process, the receiver has to pick up the carrier image, extract the parameters *brightness_level*, *brightness_mode*, and *upperbound_intensity* out of it, then select the carrier bytes based on the following formula: *UNCOVER(byte[i]- brightness_level)* IF *byte[i] – brightness_level < upperbound_intensity*. Then, for every selected carrier byte, its three LSBs have to be extracted to eventually build a long string of bits conveying the sender's original secret message M.

## 5. Experiments & Results

As a proof of concept, a simulation software was built using MS Visual C# 4.0 and MS Visual Studio 2012 under the MS .Net Framework 4.0. The software is codenamed GhostBit and it is capable of covering and uncovering secret data using the proposed method and algorithm. GhostBit has two parts, one for cover/uncovering textual data such as plaintext messages, and one for covering/uncovering generic binary data such as image files, audio files, PDF documents, and executable programs. Figure 1 shows the main GUI interface of GhostBit.

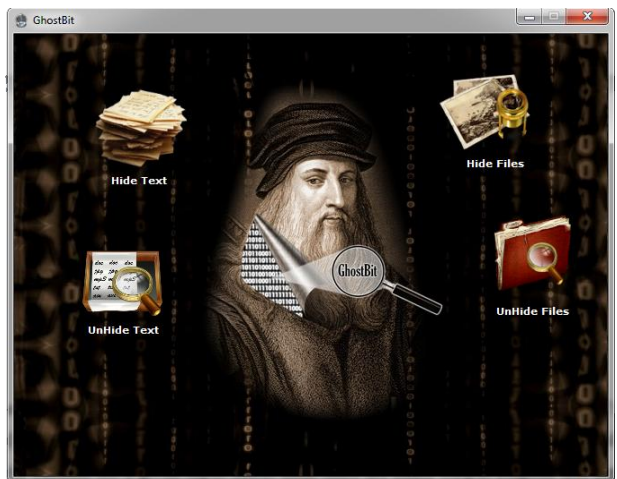

**Figure 1.** GhostBit Main GUI

The secret message is an extract from wikipedia about "Eilean Donan Island" [17] and it is equal to:

"*Eilean Donan (Scottish Gaelic: Eilean Donnain) is a small island in Loch Duich in the western Highlands of Scotland. It is connected to the mainland by a footbridge and lies about half a mile from the village of Dornie. Eilean Donan (which means simply island of Donnan) is named after Donnan of Eigg, a Celtic saint martyred in 617. Donnan is said to have established a church on the island, though no trace of this remains. The island is dominated by a picturesque castle which is familiar from many photographs and appearances in film and television. The castle was founded in the thirteenth century, but was destroyed in the eighteenth century. The present buildings are the result of twentieth-century reconstruction. Eilean Donan Castle is the home of the Clan Macrae. Eilean Donan is part of the Kintail National Scenic Area, one of 40 in Scotland. In 2001, the island had a population of just one person.*"

The carrier image is an uncompressed 24-bit BMP image with 1024x768 resolution. The parameters *brightness_level*, *brightness_mode*, and *upperbound_intensity* are predefined as *brightness_level=40*, *brightness_mode=7* (7 indicates that the bytes of three color channels are to have their brightness adjusted), and *upperbound_intensity=100* (100 indicates that the secret data are to be hidden in the bytes whose intensities are below 100). These parameters are transferred along with the carrier image hidden in the last two pixels whose zero-based coordinates are (1022,767) and (1023,767). Figure 2 shows the GUI of the covering process being executed. Figure 3 is the original image before covering the secret message into it; whereas, Figure 4 is the carrier image containing the secret message. Obviously, the brightness of Figure 4 is greater than of Figure 3 by 40 degrees (value of the parameter *brightness_level*).

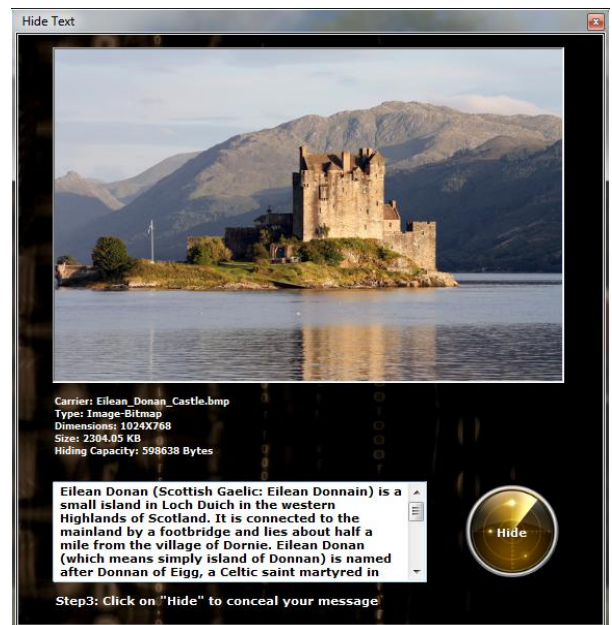

**Figure 2.** GUI of the covering process being executed

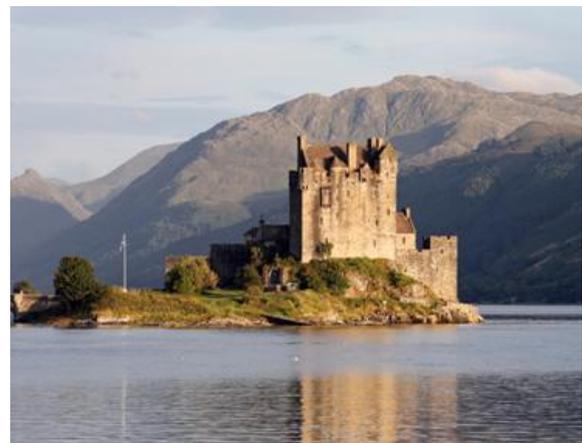

**Figure 3.** Original Image before the covering process

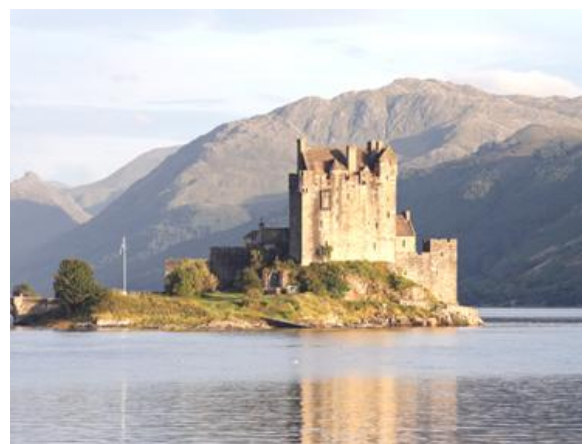

**Figure 4.** Carrier Image after the covering process

Finally, testing the uncovering process proved that the proposed algorithm is valid as it managed to recover the secret message out of the carrier image. Figure 5 shows this process.

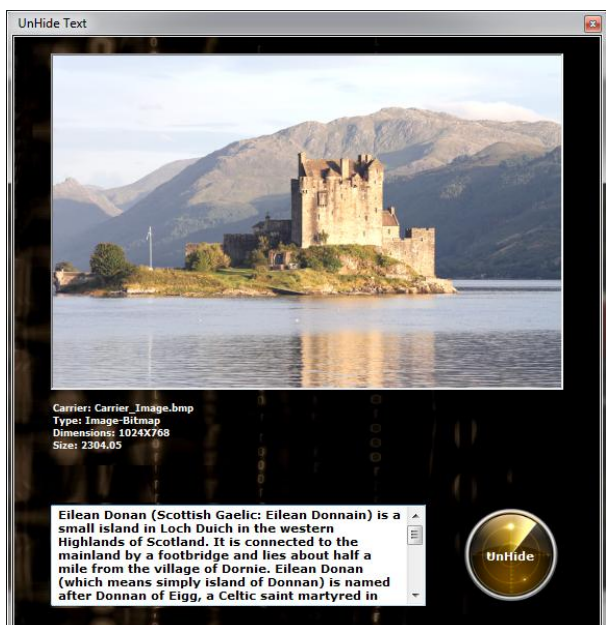

**Figure 5.** GUI of the uncovering process being executed

## 6. Conclusions

This paper proposed a novel steganography method based on adjusting the brightness of the carrier image after covering the secret data into the three LSBs of its bytes. In effect, these bytes are not selected in sequence, but in such a way that preserves the integrity of the covered data without destroying them after restoring back the brightness of the carrier image. The simulation conducted proved that the proposed algorithm is valid and correct. As a result, changing the brightness of the carrier image, prior to sending it to the intended recipient, makes the covered data irrecoverable by third parties and impossible to discover their presence, unless first, the brightness level of the carrier image is reverted back to its original state. Moreover, it is by choosing some and not all of the color channels to increase their brightness, reverse-engineering the covering algorithm would be quite tricky and ambiguous for eavesdroppers, misleading them from the true location of the covert data.

## 7. Future Work

As future work, other image processing techniques are to be investigated such as adjusting the contrast or the gamma level of the carrier image, giving the communicating parties more options to parameterize their secret communication. Furthermore, the proposed method is to be studied to see how it can be applied on carrier files other than images such as audio files. A prospective technique could be adjusting the volume of the audio file after the completion of the covering process, while not destroying the carrier audio file or the hidden secret data.

## Acknowledgments

This research was funded by the Lebanese Association for Computational Sciences (LACSC), Beirut, Lebanon, under the "Stealthy Steganography Research Project – SSRP2012".

## References

[1] Peter Wayner, "Disappearing cryptography: information hiding: steganography & watermarking", 3rd Edition, Morgan Kaufmann Publishers, 2009.

[2] Fabien A. P. Petitcolas, Ross J. Anderson and Markus G.Kuhn, "Information Hiding - A Survey", Proceedings of the IEEE, special issue on protection of multimedia content, vol. 87, no.7, pp.1062-1078, 1999.

[3] Eric Cole, "Hiding in Plain Sight: Steganography and the Art of Covert Communication", Wiley Publishing, 2003.

[4] Rafael C. Gonzalez, Richard E. Woods, "Digital Image Processing", 3rd edition, Prentice Hall, 2007.

[5] Frank Shih, "Digital Watermarking and Steganography: Fundamentals and Techniques", CRC Press, 2007.

[6] Johnson, N. F. and Jajodia, S., "Exploring steganography: Seeing the unseen", Computer Journal, vol. 31, no.2, pp.26–34, 1998.

[7] W. Bender, D. Gruhl, N. Morimoto, A. Lu, "Techniques for data hiding IBM Systems Journal", vol. 35, no 3, pp. 313-336, 1996.

[8] J. R. Smith and B. O. Comisky, "Modulation and information hiding in images," in information hiding, first international workshop, Germany: Springer-Verlag, vol. 1174, pp. 207–226, 1996.

[9] B. Pfitzmann, "Information hiding terminology", in Information Hiding, First International Workshop, vol. 1174, pp. 347–350, Springer, 1996.

[10] Tovée, Martin J., "An introduction to the visual system", Cambridge University Press, 2008.

[11] R. Anderson, F. Petitcolas, "On the limits of steganography," IEEE Journal on Selected Areas in Communications, vol. 16, 1998.

[12] Ross J. Anderson, "Information hiding: 1st international workshop", volume 1174 of Lecture Notes in Computer Science, Isaac Newton Institute, Springer-Verlag, Germany, 1996.

[13] T. Zhang and X. Ping, "A Fast and Effective Steganalytic Technique against JSteg-like Algorithms", Proceedings of the 8th ACM Symposium, Applied Computing, ACM Press, 2003.

[14] Eiji Kawaguchi and Richard O. Eason, "Principle and applications of BPCS-Steganography", Proceedings of SPIE: Multimedia Systems and Applications, vol.35, no.28, pp.464-473, 1998.

[15] Maria Petrou, Costas Petrou, "Image Processing: The Fundamentals", 2nd edition, Wiley, 2010.

[16] J.R. Parker, "Algorithms for Image Processing and Computer Vision", 2nd edition, Wiley, 2010.

[17] Wikipedia article entitled Eilean Donan, URL: http://en.wikipedia.org/wiki/Eilean_Donan